\documentclass[reprint,nofootinbib,amsmath,amssymb,aps]{revtex4-1}

\usepackage{graphicx}
\usepackage{dcolumn}
\usepackage{bm}
\usepackage{hyperref}
\usepackage{titlesec}
\usepackage{color}
\titleformat*{\section}   {\centering\normalsize\bfseries}
\titleformat*{\subsection}{\centering\small\bfseries}

\titlespacing*{\section}   {0pt}{1\baselineskip}{0.01\textwidth}
\titlespacing*{\subsection}{0pt}{1\baselineskip}{0.01\textwidth}

\renewcommand{\figurename}{Fig.}
\renewcommand{\tablename}{Table.}
\def\arraystretch{1.5}

\newcommand{\ep}{e^{+}e^{-}}
\newcommand{\mumu}{\mu^{+}\mu^{-}}
\newcommand{\GeV}{\rm ~GeV}

\newcommand{\invfb}{\rm ~fb^{-1}}
\newcommand{\pref}[1]{~\ref{#1}}
\newcommand{\pcite}[1]{~\cite{#1}}
\usepackage{ mathrsfs }
\usepackage{ amsmath }
\usepackage{ amssymb }
\begin{document}

\preprint{APS/123-QED}

\title{Feasibility of a minimum bias analysis of ${\bf e^+e^-\rightarrow ZH\rightarrow q\bar{q}+X}$ at a ${\bf 250}$~GeV ILC}
\thanks{Talk presented at the International Workshop on Future Linear Colliders (LCWS13)~\cite{yhaddad:LCWS13Online}, Tokyo, Japan, 11-15 November 2013.}%

\author{Yacine Haddad}
\email{yhaddad@cern.ch}
\affiliation{%
  Laboratoire Leprince-Ringuet (LLR), \'Ecole polytechnique, CNRS-IN2P3.
}

\date{\today}

\begin{abstract}
  
  The precision measurements of the Higgs properties is crucial for a better understanding of electroweak symmetry breaking. 
  It can be first achieved at the ILC\pcite{Behnke2013} at $\sqrt{s}=250$ via the Higgs-strahlung production process $e^+e^-\rightarrow ZH$.  The hadronic decay mode $Z\rightarrow \bar{q}q$ constitutes more than $65\%$ of the total, a factor 10 more than $Z\rightarrow \mu\mu$.  An analysis based solely on the $Z$ jet pair information could thus lead to a high statistics and provide a minimum biased Higgs sample. 
  A study of the feasibility of such analysis is shown here, based on $e^+e^-$ simulated collisions at $250\rm ~GeV$ in center of mass, for the equivalent integrated luminosity of $500\rm ~fb^{-1}$ and using a fast simulation of the ILD detector.
  
\end{abstract}
\maketitle

\section{\label{sec:Introduction}Introduction}

\begin{figure}[!h]
  \centering
  \includegraphics[width=0.35\textwidth]{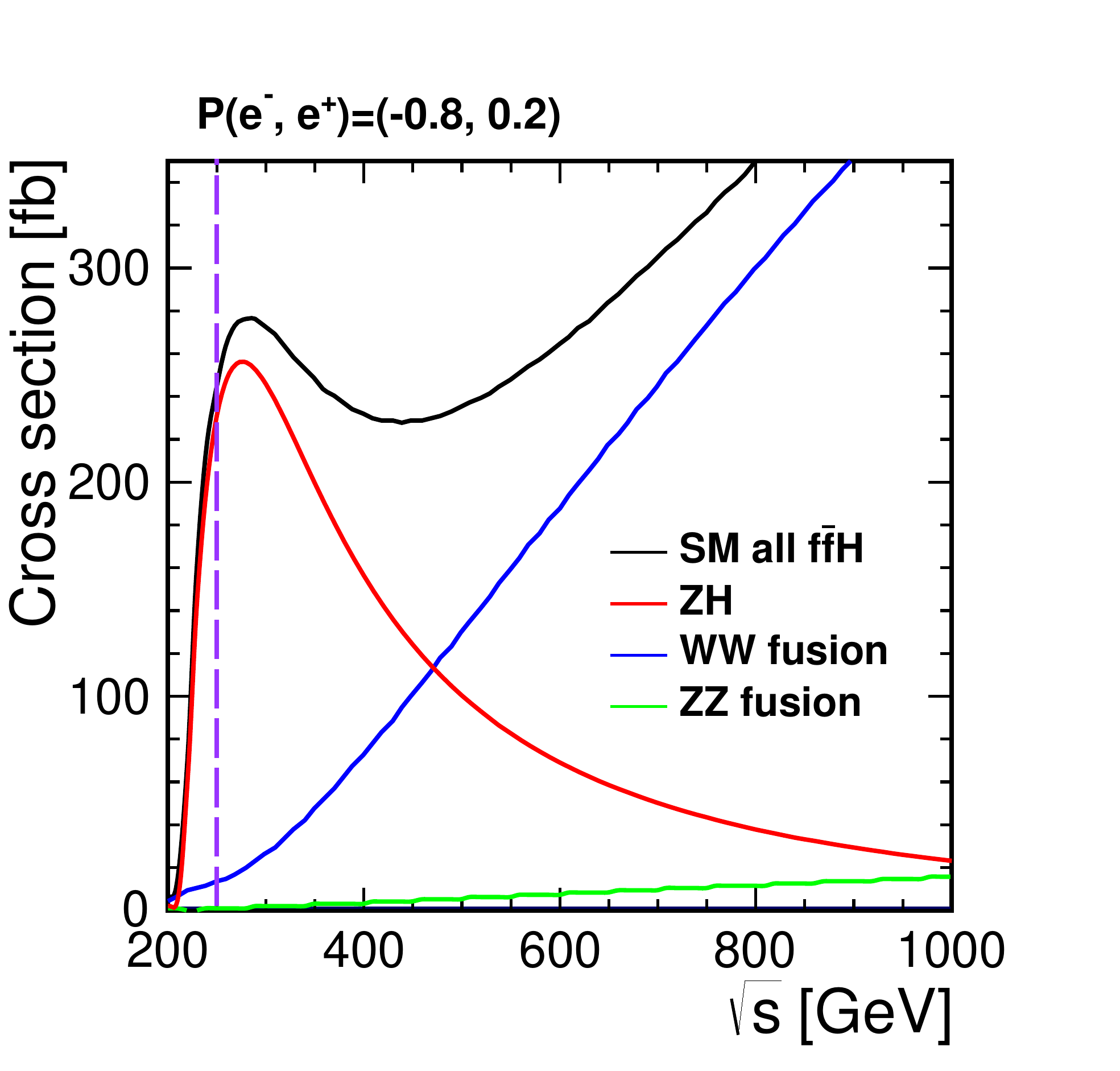}
  \caption{Higgs boson production cross section as function of the $\ep$ center of mass energy. The purple dashed line represents the ILC operating at a center-of-mass energy of $\sqrt{s}=250\GeV$}
  \label{fig:xsec}
\end{figure}

\begin{figure}[!h]
  \centering
  \includegraphics[width=0.25\textwidth]{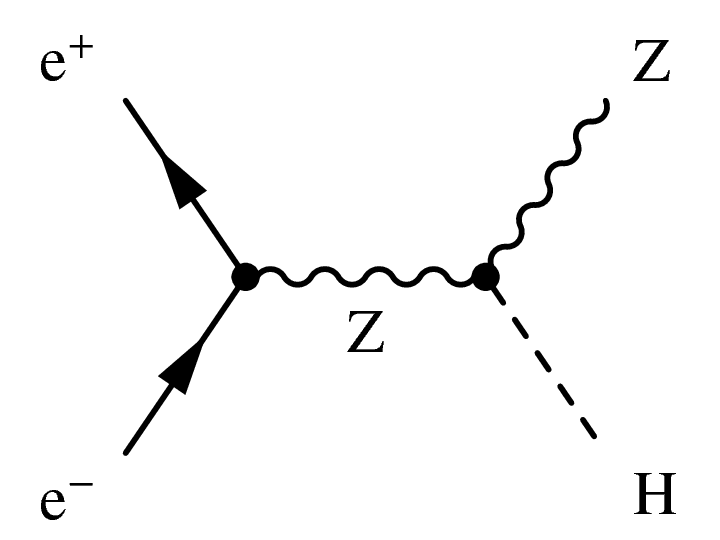}
  \caption{Leading order Feyman diagram of the Higgs boson production via Higgs-strahlung process.}   
  \label{fig:feynm}
\end{figure}

The measurement of the Higgs boson properties at the future $\ep$ collider ILC\pcite{Behnke2013} can be achieved via the known Higgs-stralung process (Fig.\pref{fig:feynm}). The $\ep\rightarrow ZH$ represents the largest production cross section for a center-of-mass (c.o.m.) energy of $\sqrt{s}=250\GeV$ assuming a Higgs mass of $125\GeV$ (Fig.\pref{fig:xsec}).

The well defined center-of-mass energy of collision allows to perform analyses independent from the Higgs boson decaying products. Indeed, the identification of the $ZH$ signature can be made by reconstructing the $Z$ boson only, and selecting the proper recoiling mass against its decay products. This approach allows to measure the Higgs branching ratio and Higgs production cross section independently from Higgs decay modes, including invisible Higgs ones, such as $H\rightarrow ZZ \rightarrow \nu\bar{\nu}\nu\bar{\nu}$. 


Previous studies~\cite{Li2012} have addressed the case where the $Z$ boson decays to $\ep$ or $\mumu$ pairs. For these leptonic channels, the energy and the momentum can be precisely measured with a high tracking detector performance such as the ones proposed for the ILC detectors (SiD and ILD~\cite{Behnke2013}) aiming at momentum resolution of about $\sigma_{(1/p_{t})}\sim 2\times10^{-5}\GeV^{-1}$. A precision of about $2.6\%$ on the $ZH$ cross section can be achieved at ILC~\cite{Li2012}. 

However, analyses considering only the leptonic channels, are limited by the statistical precision due to the small branching ratio of $Z \rightarrow \mumu$ and $Z\rightarrow \ep$ process ($\sim 3.3\%$). Thus the idea is to extend this analysis to hadronic decay mode of the $Z$ boson\footnote{The case of $Z$ boson decaying into pair of $\tau$ leptons, which can fake the hadronic decay, is not considered here.}, $Z\rightarrow q\bar{q}$, which represent the large branching ratio of about $\sim 70\%$. The challenge is greater since it depends on the jet clustering algorithm which may introduce confusion in the recombination process for hadronic decays of the Higgs boson (and $H\rightarrow \tau\tau$). Thus different reconstruction efficiency are expected for each decay mode of the Higgs boson.

The following study is based on a fast simulation. 
A smearing of the four momentum of the stable particles takes care of the expected performance of the ILD detector.
The jets are then reconstructed using standard $\ep$ jet clustering. The $Z$ boson jet pair is selected by having mass consistent with $Z$ boson. 
In order to reduce the background contamination, an event selection based on s Boosted Decision Tree (BDT) \cite{Therhaag2009} is exploited. The first preliminary results on this $ZH$ analysis are here shown and commented upon. 

\section{\label{sec:SimTools}Simulation tools}
\subsection{Event generation}

All events were generated using \verb;WHIZARD;~1.95 Monte Carlo (MC) event generator\pcite{Kilian:2007gr} configured with ILC beam parameters, taking into account beamstrahlung and initial state radiation (ISR) photons. \verb;PYTHIA;~6.4\pcite{Sjostrand:2006za} insured the final state QED and QCD parton showering, fragmentation and decay providing final-state observable particles. 


Both signal and background events were produced at a c.o.m energy of $\sqrt{s}=250\GeV$ for a total integrated luminosity of $500\invfb$ and a Higgs mass of $125\GeV$. The $ZH$ process accounts for $95\%$ of the total fermionic $f\bar{f}H$ cross section for the chosen Higgs mass (Fig.~\ref{fig:xsec}). The SM processes $\ep\rightarrow WW$ and $\ep\rightarrow ZZ$ decaying hadronically, leptonicaly or semi-leptonically are the main backgrounds. The Table.\pref{tab:samples} summarizes the statistics of the generated samples, signal and background, as well as the corresponding cross section and event weight\footnote{The event weight for a given process with a production cross section of $\sigma_{proc}$ is defined as $w=N_{gen}/\mathscr{L}\sigma_{proc}$ for a luminosity $\mathscr{L}$ and a  number of generated event $N_{gen}$.}. 


\begin{table}[!h]
  \begin{tabular}{lccc} 
    \hline\hline
    Process & $N_{events}$ & $\sigma~(fb)$   & weight \\  & & $e^-_Le^+_R$&($\mathscr{L}= 500~fb^{-1}$) \\ \hline
    \textcolor{red}{$ZH\rightarrow q\bar{q}+X$}    &  \textcolor{red}{120409} & \textcolor{red}{346.01}     &  \textcolor{red}{1.41}  \\ 
    $WW\rightarrow q\bar{q}q'\bar{q'} $            &  321376 & 18781.00   &  60.48 \\ 
    $WW\rightarrow q\bar{q}l\nu $                  &  181533 & 14874.30   &  52.30 \\
    $ZZ\rightarrow q\bar{q}q'\bar{q'} $            &  120088 & 1422.14    &  4.45  \\
    $ZZ\rightarrow q\bar{q}ll $                    &  178900 & 1402.06    &  6.46  \\
    \hline
  \end{tabular}
  \caption{Statistic of the generated samples as well as the corresponding event weight.}
  \label{tab:samples}
\end{table}


Here only the $e^-_Le^+_R$ beam polarization was considered. It yields the largest production cross section and is the most challenging for the background suppression.

\subsection{Generic fast simulation}



The response of the ILD detector is modeled by a fast Monte-Carlo simulation (fast-MC), smearing each particle four-momentum according to the expected precision. A Particle Flow Algorithm (PFA) analysis is then used, in which charged particles are measured in the tracker (ignoring their calorimeter deposits) while only neutral particles (photon, neutral hadron) are measured by the calorimeters. Neutrinos are ignored.

\begin{figure}[!h]
  \centering
  \includegraphics[width=0.35\textwidth]{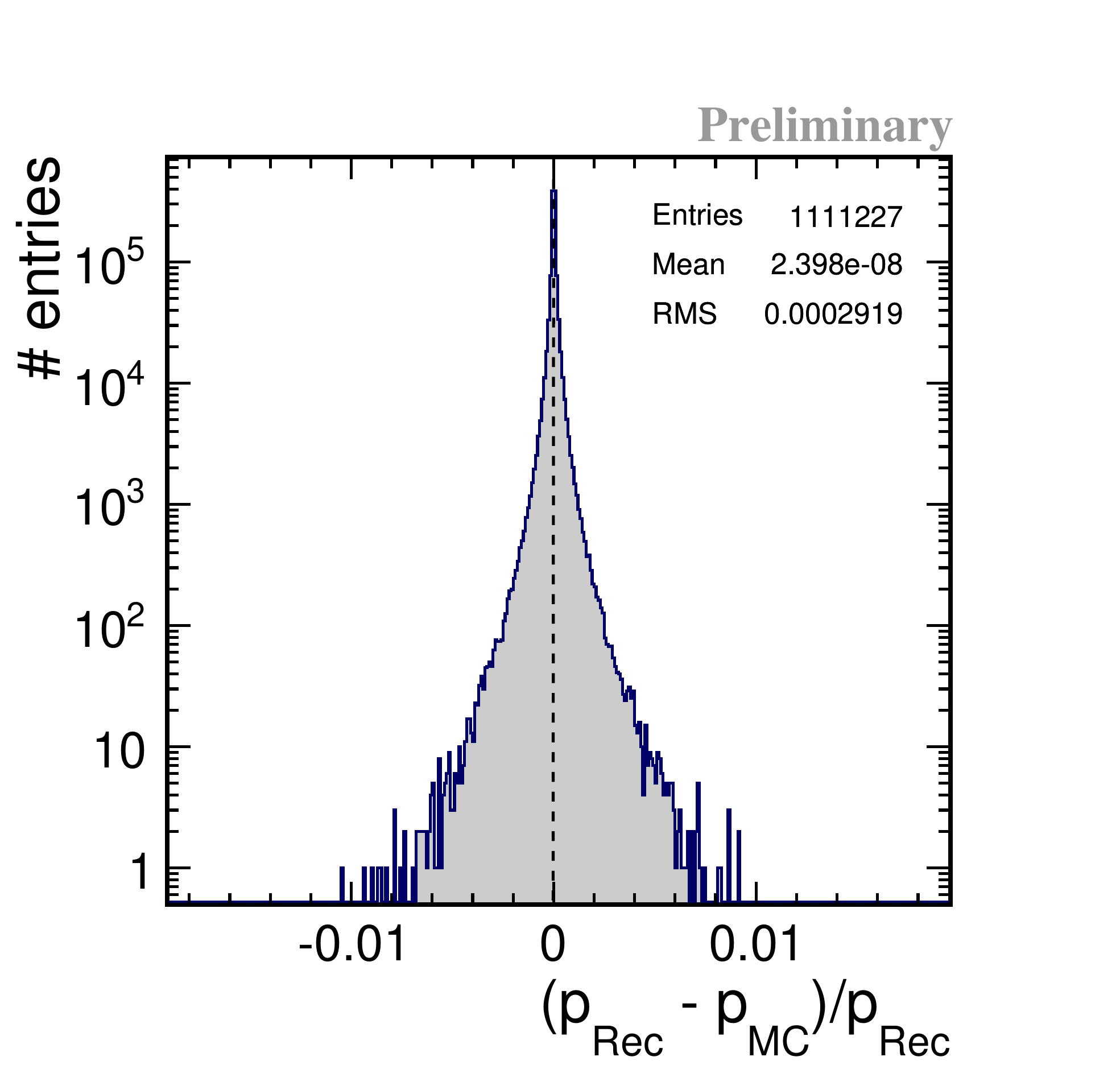}
  \caption{Difference between the momentum of charged particle before and after smearing. The charged particles are well measured with respect to the performance of the tracker resolution.}
  \label{fig:charged}
\end{figure}

The charged particle transverse momentum is smeared by a Gaussian of standard deviation: 
\begin{equation}
  \sigma_{1/p_T} \approx 2\times 10^{-5}~\text{GeV}^{-1}
  \label{eq:1}
\end{equation}
The Fig.~\ref{fig:charged} represent the relative dispersion of charged particle momentum. 
For photons ($\gamma$) and neutral hadrons ($h^0$) the energy is smeared by the expected calorimetric performance which correspond roughly to 
\begin{eqnarray}
  \gamma:&& ~~\sigma_{E}/E \sim 0.01 \oplus 0.1/\sqrt(E ~[\GeV])\nonumber\\
  h^0:   &&  ~~\sigma_{E}/E \sim 0.1 \oplus 0.5/\sqrt(E ~[\GeV]) 
  \label{eq:xdef}
\end{eqnarray}

Each reconstructed particle is considered as Particle Flow Object (PFO). Only PFO with  $|\eta| <2.66$ and $p_t>0.5~\rm GeV$ are kept to mimic the acceptance of the ILD detector. 

\section{\label{sec:Analysis}Analysis Strategy}
\subsection{Jet Reconstruction}

\begin{figure}[!h]
  \centering
  \includegraphics[width=0.35\textwidth]{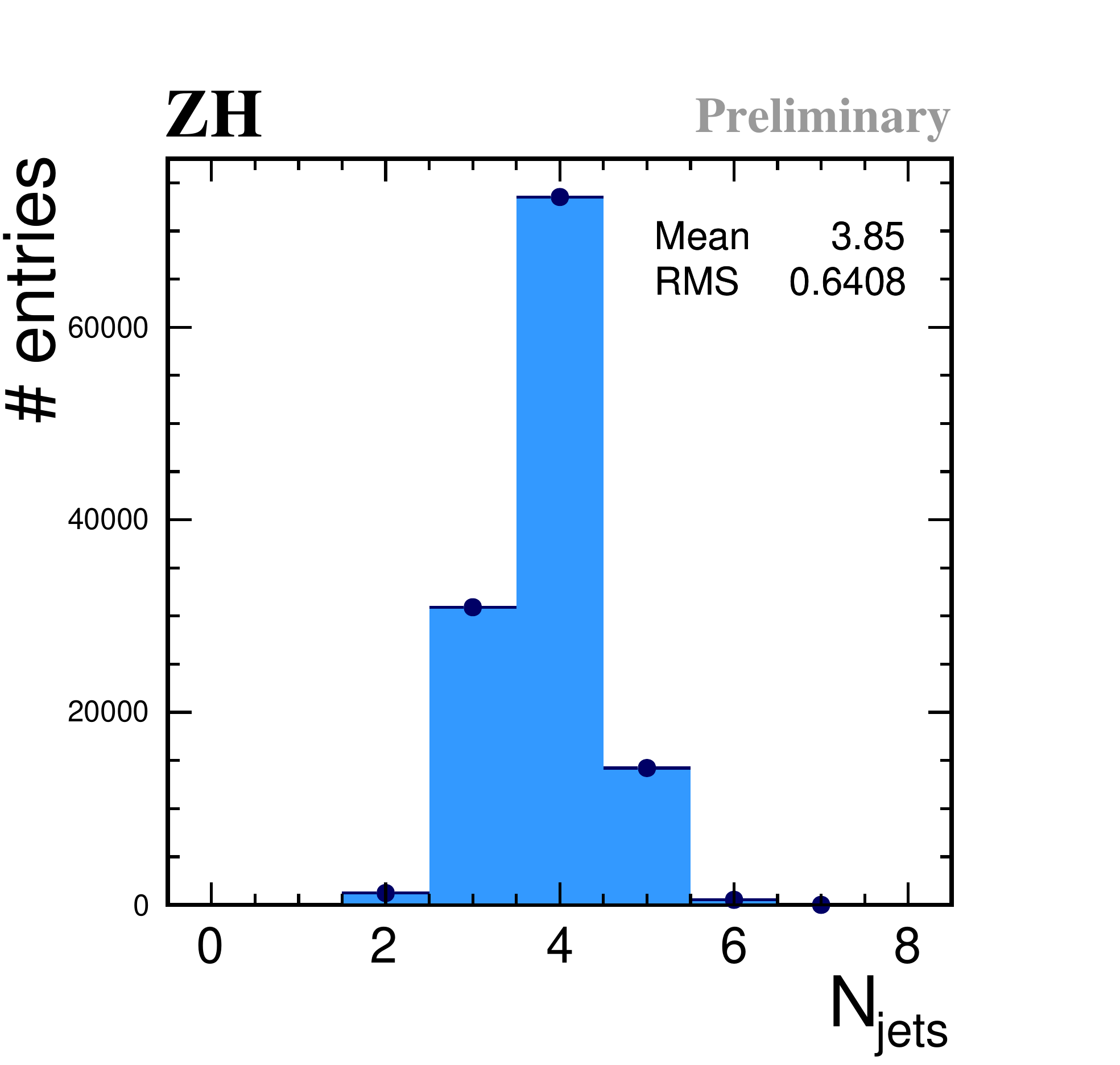}
  \caption{ Distribution of number of jet for signal events after Durham-$k_{t}$ jet clustering, with a fixed $y_{cut}=0.01$}
  \label{fig:njets}
\end{figure}

The fragmentation products of the hadronic system were clustered using the Durham-$k_t$ algorithm\pcite{Catani1992291,raey} implemented in \verb;FastJet;\pcite{Cacciari:2011ma}. The Durham-$k_{t}$ algorithm has only one parameter $y_{cut}$ fixed here at $y_{cut}=0.01$ without restricting the number of reconstructed jets. The particles are combined using the energy combination scheme where the four-vector of particle are summed. The Fig.\pref{fig:njets} shows the distribution of number of jets of the signal. A peak around $N_{\rm jets}\sim 4$ is observed for the signal as expected from the predominantly hadronic decays of the $Z$ and Higgs bosons.

All the possible jet pair combination is calculated. The pair matching at the best a $Z$ boson is selected using the $\chi2 = (m_{jj}-m_{z})^2/\sigma_{z}^2$ criterion. 

\subsection{Boosted Decision Tree (BDT) based selection}

The separation of the signal and background event is performed by a Boosted Decision Tree (BDT) as implemented in the \verb;TMVA;~\cite{Therhaag2009} package. 

The set of variables used for the event selection as input of the BDT training is first restricted to the ones relative to the selected jet pair, such as $E$, $E_{visible}$, ${\rm cos}\theta_{jj}$ and $\chi^2$(from di-jet pairing). The \textit{Acolinearity} and \textit{Acoplanarity}, defined by $A_{acol} = \cos^{-1}(\vec{p_{1}}\cdot\vec{p_{2}}/|\vec{p_{1}}||\vec{p_{2}}|)$ and $\Delta \phi = |\phi_1 - \phi_2|$ respectively, are also used, where $\phi_i$ is the angle of a jet $i$ in the transverse plane with respect to the beam axis.

The di-jet only variables are not sufficient for a good background rejection. Two additional \textit{Event Shape} variables defined as:  
\begin{itemize}\setlength{\itemsep}{0.0\baselineskip}
\item Thrust\cite{Brandt1964}:
  $\tau = 1 - \max_{\vec{n}}\left(\frac{\sum_i|\vec{p}_{i}\cdot \vec{n}|}{\sum |\vec{p}_i|}\right)$
\item Sphericity\cite{Brodsky1995}:
  $S = \frac{3}{2}\min\left(\frac{\sum \vec{p}^2_{L}}{\sum \vec{p}^2}\right)$
\end{itemize}
are included to the selection set. $\vec{n}$ is a unit vector.


Additional variables coming from the jet clustering, called transition parameters, are also used. A jet transition parameter $y_{k,k+1}$ is defined are the value of $y_{cut}$ in which the event flip from $(k+1)$-jet to $(k)$-jet configuration\footnote{Using this value, a selection of $y_{cut}$ can be made such that the event is resolved into the required number of jets.}. $y_{23}$, $y_{34}$ and $y_{45}$ are appended to BDT input variable set. Note that the recoil mass variable is not used in the BDT training to minimize as much as possible the selection bias. 

The selection is done in two steps. First, the event are divided in four categories corresponding to each considered background (WW/ZZ decaying whether hadronically or semi-leptonically). The training of the BDT is applied to recognize the signal from each background category. Four output variables are then obtained. The second step, consists then in a cut base selection on the four BDT-variables to reduce each background category. Events surviving the BDT selection are used for the signal strength estimation.   

\begin{figure}[!h]
  \centering
  \includegraphics[width=0.45\textwidth]{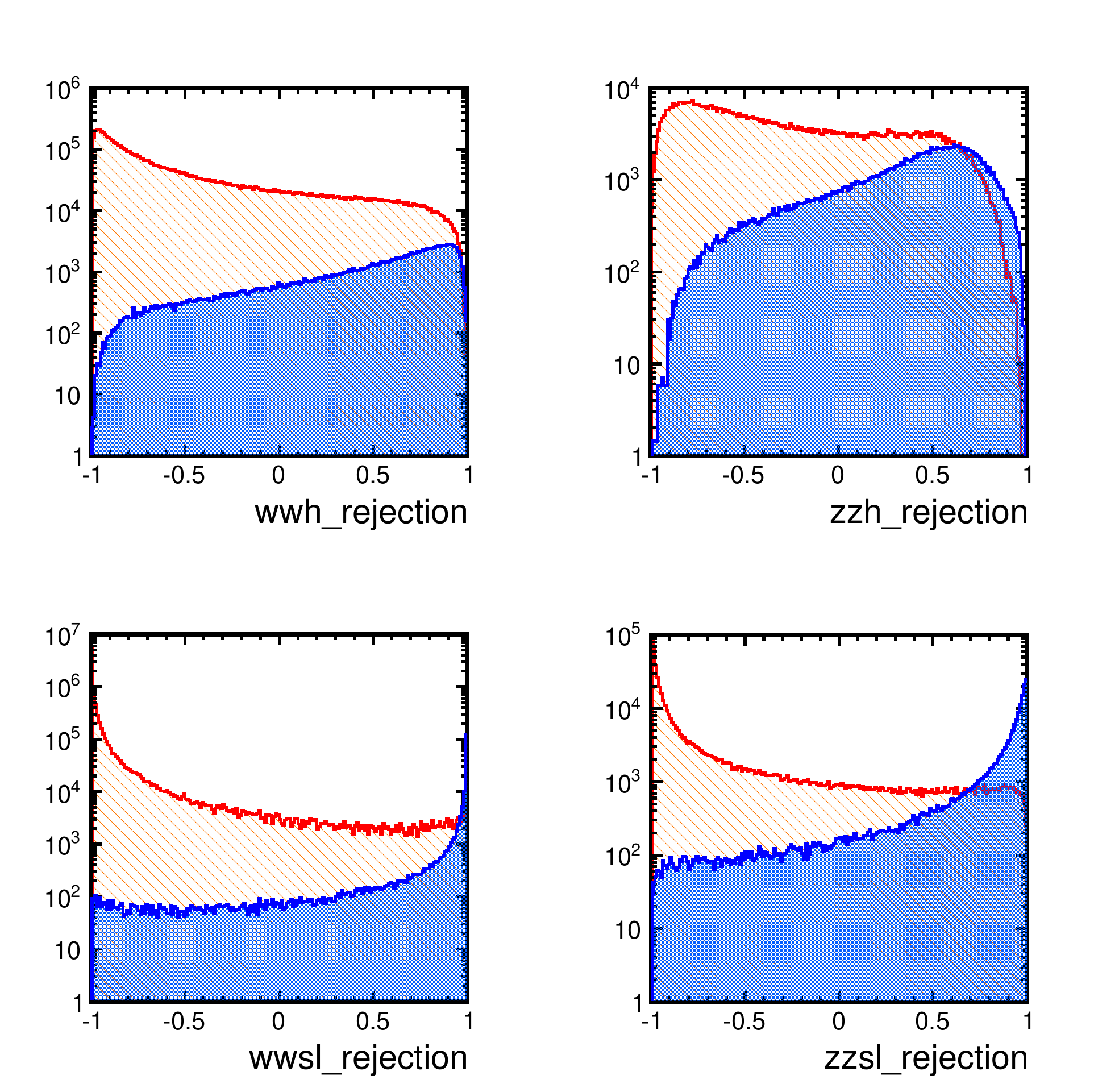}
  \caption{Distribution of the BDT output variable for the different background categories. The red and blue lines represents background and the signal respectively.}
  \label{fig:bdtoutput}
\end{figure}

Several options of BDT algorithm are available in the \verb;TMVA; package, we choose the BDT-Gradient (BDTG) which shows a good signal-background separation. The BDT output variable ranges in $[-1,1]$ (Fig.\pref{fig:bdtoutput}).  

By varying a cut on the BDT output variable, the signal efficiency and the background rejection efficiency were determined. The signal efficiency is defined as the fraction of the event passing the cut, the background rejection efficiency is the fraction of the background event removed by the applied cut.     

\begin{figure}[!h]
  \centering
  \includegraphics[width=0.45\textwidth]{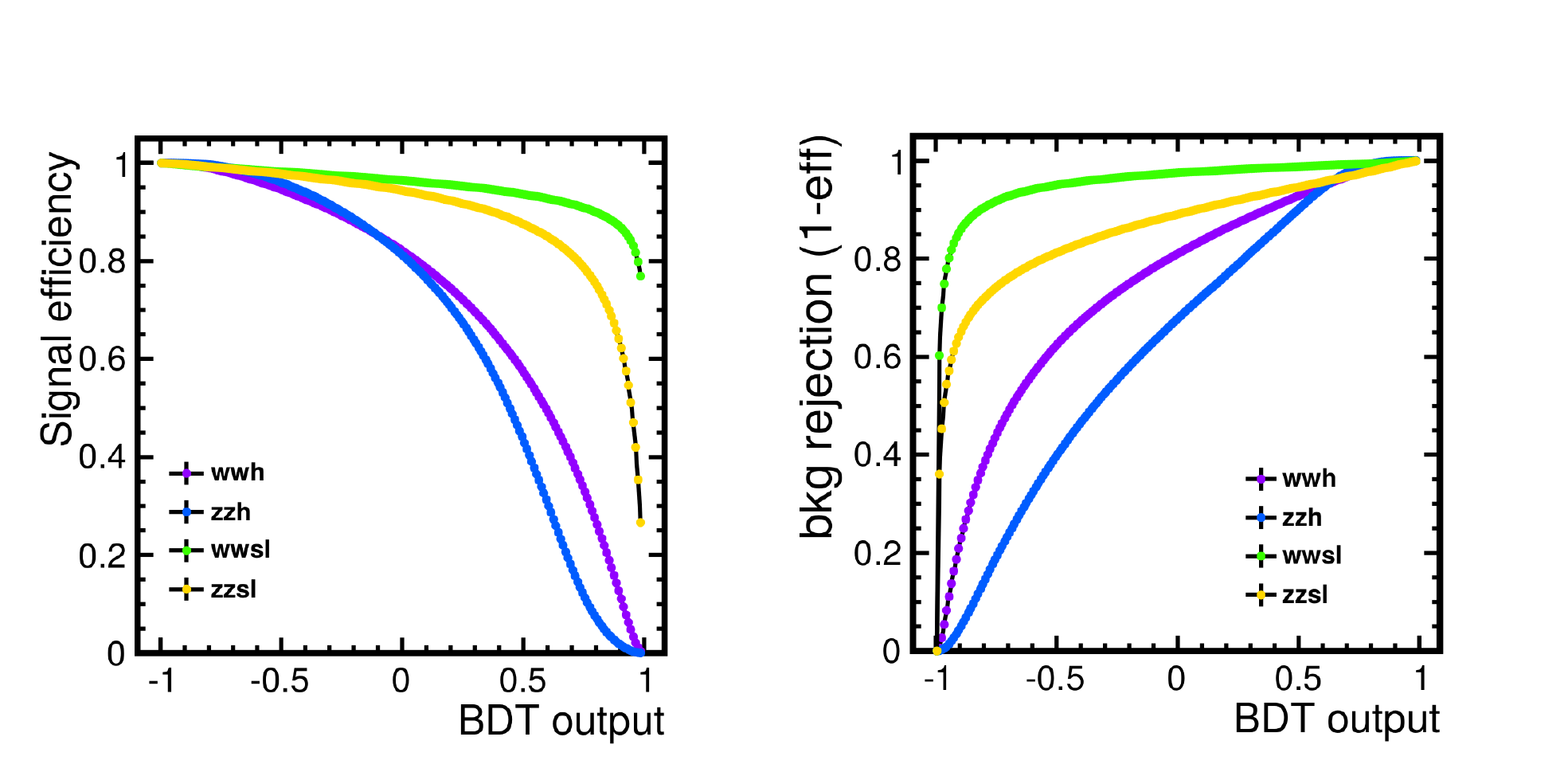}
  \caption{Signal efficiency (left ) and background rejection (right) curves of each trained category.}
  \label{fig:seleceff}
\end{figure}

An appropriate cut can then applied on BDT output variables as shown in the Table.\pref{tab:bdtcut} to insure maximum background rejection as well as a good signal efficiency. The results shown that the four backgrounds are significantly reduced with an efficiency above $94\%$, for a signal efficiency above $44\%$ (for the category $ZZ \rightarrow 2j$) and up to $98\%$(for the category $WW \rightarrow 2j+2l$) 

\begin{table}[!h]
  \begin{tabular}{lllcc} 
    \hline\hline 
    Category & &$\rm BDT_{cut}$  & \textcolor{red}{$\varepsilon_{sig}$} & $ (1-\varepsilon_{bkg})$ \\ 
    \hline
    $\rm{WW \rightarrow 4j   }$ &$\rm BDT_{output}$& $> 0.8  $ & \textcolor{red}{$58\%$} & $98.2 \%$\\ 
    $\rm{WW \rightarrow 2j+2l}$ &$\rm BDT_{output}$& $ > -0.5$ & \textcolor{red}{$96\%$} & $95   \%$\\ 
    $\rm{ZZ \rightarrow 4j   }$ &$\rm BDT_{output}$& $> 0.6  $ & \textcolor{red}{$44\%$} & $94   \%$\\ 
    $\rm{ZZ \rightarrow 2j+2l}$ &$\rm BDT_{output}$& $ > 0.6 $ & \textcolor{red}{$88\%$} & $94.5 \%$\\ 
    \hline
  \end{tabular}
  \caption{Summary of the applied cut on the BDT output for the different categories. The efficiency $\varepsilon_{sig|bkg}$ is defined as the fraction of the event passing the cuts.}
  \label{tab:bdtcut}
\end{table}

\section{\label{sec:Results}Results}

\begin{figure}[!h]
  \centering
  \includegraphics[width=0.3\textwidth]{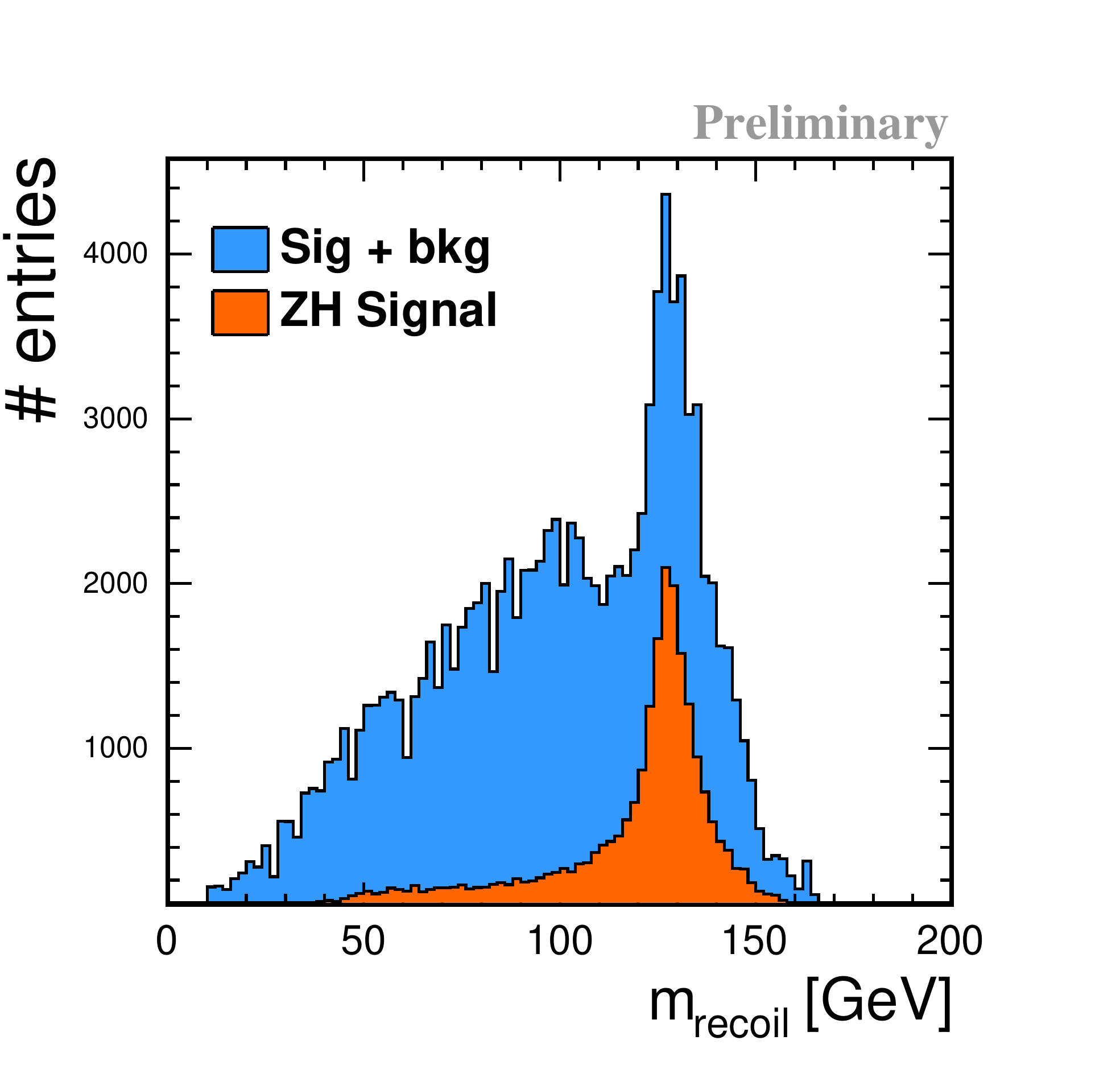}
  \caption{distribution of the reconstructed Higgs recoil mass for $ZH\rightarrow q\bar{q}X$.}
  \label{fig:results}
\end{figure}

The recoiling Higgs mass distribution after the selection of the $ZH\rightarrow q\bar{q}+X$ is shown in the Fig.\pref{fig:results}. A clear signal peak can be easily identified. 

An overall background rejection of over $98\%$ is notified (Table.\pref{tab:cutflow}), but with a poor signal efficiency of about $\sim 14\%$. This low efficiency is due to the difficulties to remove background event (mainly $WW/ZZ\rightarrow q\bar{q}q\bar{q}$) having the same topology as the signal.  

\begin{table}[!h]
  \begin{tabular}{lcccc} 
    \hline\hline
    Samples   & Before cuts   & After cuts    &efficiency \\ 
    \hline
    \textcolor{red}{$\rm{ZH }$} & \textcolor{red}{120409} & \textcolor{red}{16613}&\textcolor{red}{13.8\% }  \\ 
    $\rm{WW\rightarrow 4j} $  & 321376 & 2225 & 0.6\%   \\ 
    $\rm{WW\rightarrow 2j+2l}$ & 120088 & 246  & 0.1\%   \\ 
    $\rm{ZZ\rightarrow 4j} $  & 120088 & 3085 & 2.5\%   \\ 
    $\rm{ZZ\rightarrow 2j+2l}$ & 178900 & 967  & 0.5\%   \\ \hline
  \end{tabular}
  \caption{ Signal and background selection efficiency.}
  \label{tab:cutflow}
\end{table}

\section{Conclusion and prospective}
A method based on the multivariate technique for the analysis of the recoiling system of the $ZH$ where $Z$ decays hadronically is proposed. It uses a generic detector fast simulation, tuned on the ILD detector benchmarks. A first study based on four background categories shows a rejection of $98\%$ of the considered backgrounds. However, the present selection suffers from the large signal suppression, only $14\%$ of the signal surviving the cuts. 

Next, the $q\bar{q}$ (continuous Drell-Yan backgroung) has to be added. The $Z$ jet pair selection could be improved by introducing other criteria, such as the jets charge and Optimizing the MVA selection.

The current analysis cannot not yet conclude on the efficiency selection on each decay mode of the Higgs. Thus complementary studies are ongoing.

\bibliography{mybib}{}
\bibliographystyle{elsart-num-names}
\end{document}